\documentclass[a4paper,12pt]{article}

\usepackage[latin2]{inputenc}
\usepackage{graphicx}
\usepackage{amstext}
\usepackage[]{ntheorem}
\usepackage{fancyhdr}
\usepackage{amsfonts,dsfont}
\usepackage{natbib}
\usepackage{amsmath,amssymb}
\usepackage{t1enc}
\usepackage{authblk}

\theoremstyle{definition}

\newcommand{\PP}{\mathcal{P}}
\newcommand{\SSS}{\mathcal{S}}
\newcommand{\EE}{\mathcal{E}}

\title{Modelling transfer profits as externalities in a cooperative game-theoretic model of natural gas networks}
\author[1]{D\'{a}vid Csercsik*}
\author[2]{Franz Hubert}
\author[3,4]{Bal\'{a}zs R. Sziklai}
\author[3,5]{L\'{a}szl\'{o}~\'{A}.~K\'{o}czy}
\affil[1]{Faculty of Information Technology and Bionics, P\'{a}zm\'{a}ny P\'{e}ter Catholic University, Pr\'{a}ter utca 50/A, H-1083 Budapest, Hungary, \emph{csercsik@itk.ppke.hu}}
\affil[2]{School of Business and Economics, Humboldt-Universit\"{a}t zu Berlin, Spandauer Stra{\ss}e 1, D-10178 Berlin, Germany}
\affil[3]{Centre for Economic and Regional Studies, Hungarian Academy of Sciences, T\'{o}th K\'{a}lm\'{a}n utca 4, H-1097 Budapest, Hungary}
\affil[4]{Faculty of Economics, Corvinus University Budapest, F\H{o}v\'{a}m t\'{e}r 8, H-1093, Budapest, Hungary}
\affil[5]{Faculty of Economic and Social Sciences, Budapest University of Technology and Economics\\
              Magyar tud\'{o}sok k\"{o}r\'{u}tja 2. H-1117 Budapest, Hungary}
\date{}                     
\begin{document}
\maketitle
\begin{abstract}
Existing cooperative game theoretic studies of bargaining power in
gas pipeline systems are based on the so called characteristic function form
(CFF). This approach is potentially misleading if some pipelines fall
under regulated third party access (TPA). TPA, which is by now the norm
in the EU, obliges the owner of a pipeline to transport gas for others,
provided they pay a regulated transport fee. From a game theoretic
perspective, this institutional setting creates so called
"externalities," the description of which requires partition function form
(PFF) games.  In this paper we propose a method to compute payoffs,
reflecting the power structure,  for a pipeline system with regulated
TPA. The method is based on an iterative flow mechanism to determine
gas flows and transport fees for individual players and uses the
recursive core and the minimal claim function to convert the PPF game
back into a CFF game, which can be solved by standard methods. We
illustrate the approach with a simple stylized numerical example of the gas network in Central Eastern Europe with a focus on Ukraine's power index as a major transit country.
\textbf{Cite as}: Csercsik et. al. Energy Economics, 2019 \\ https://doi.org/10.1016/j.eneco.2019.01.013

\end{abstract}


 \section{Introduction}\label{Introduction}

Markets for natural gas depend on long-lived and idiosyncratic, hence, sunk, investments in technical infrastructure. In particular, pipelines create large quasi rents, and the power to appropriate these rents has attracted substantial academic interest. The literature has pursued two different approaches using tools from non-cooperative and cooperative game theory, respectively.

 The non-cooperative approach starts with a description of how players interact: their strategy spaces. After solving for the Nash-equilibrium strategies, the individual payoffs and the overall welfare are calculated. The former reveal the market power of a player, often measured as a mark-up on marginal cost, and the latter indicates the efficiency of cooperation. The literature has developed  disaggregated models with a large number of players.\footnote{For multilevel oligopoly models of the European/Eurasian network, see, among others, \citet{deWolf1997stochastic}, \citet{gurkan1999sample} \citet{boots2004trading}, and \citet{Abada2013}.  \citet{breton2001equilibria}  compare the Cournot and Stackelberg equilibria in this market. On the North American natural gas system,  \citet{gabriel2004nash,gabriel2005large,gabriel2005mixed} provides a series of papers among which they use computational and (Nash-Cournot) equilibrium methods.} However, in order to keep the analysis manageable, highly restrictive assumptions on the strategy space of the players (usually linear prices or quantities) and on the sequencing of moves, hence, the ability to commit, have been adopted. These assumptions are clearly at odds with the complexity of negotiations and the widespread use of comprehensive price-quantity contracts, which we observe in the European gas markets.\footnote{These contracts stipulate the total quantity of gas produced and/or shipped and the total payment. The quantity can be set to ensure efficiency, and the payment determines how the surplus is shared. Such contracts, also referred to as 'take-or-pay contracts' in the gas industry, can avoid the inefficiencies of double marginalization along the vertical supply chain: production/transport/consumption (see \cite{ECS2007-Regulation} for more details). }  Moreover, the results on market power and the alleged inefficiency of the market are largely driven exactly by these counter-factual assumptions (see \citet{hubert2014access}).

 The cooperative approach, in contrast, rests on the assumption that rational players are able to coordinate their activities efficiently. Gas producers, pipeline owners, and gas consumers will trade the efficient amounts of gas, but the payments to producers and pipeline owners, hence, the sharing of economic surplus between the participants, will reflect the players' power in the network. Intuitively, the power of a player will reflect his control of important infrastructure, such as gas fields or pipelines or facilities (e.g.\ power plants), and on the difficulty of replacing his role in a given network. The primitive of cooperative game theory is the description of the gains from cooperation. It is obtained as follows: take an arbitrary (sub)group of players, usually referred to as \emph{coalition}, and calculate how much surplus they can generate for themselves if they use the means at their disposal in the best possible way. The result of this optimization problem is called the value of the coalition. For a gas network, it requires the following: first, finding the subnetwork (gas producing fields, pipelines, and gas consuming appliances) that the coalition has access to, and second, determining the optimal usage of the subnetwork and the resulting surplus.\footnote{When setting up this optimization problem one can also account  for limits in the ability to coordinate (\citet{roson2015bargaining}).}  By repeating this thought experiment for all possible coalitions we obtain a complete description of the gains from cooperation in the given environment, the so-called value function. Finally, the game is solved by determining how the players share the gains resulting from cooperation by assigning a payoff to every player. A player's share in the total surplus can be interpreted as his power index. There is no consensus in cooperative game theory on which theoretical solution is best; the applied literature has explored the Shapley Value \citep{Shapley1953} as well as the nucleolus \citep{Schmeidler1969}.

 A number of studies have used the cooperative approach to study how pipeline investments and access rights affect the power relation in the European/Eurasian gas network. The literature explores different venues for the treatment of investments and access rights and assumes that the players, usually interpreted as countries or the dominant firm in a region, can make efficient use of the existing infrastructure.  \citet{hubert2011investment} study how options to invest in new pipelines or upgrading existing ones affect the power structure between Russia and major transit countries towards North-Western Europe.  \citet{hubert2008dynamic} consider the same geographical region and ask whether dynamic strategies in repeated games can support efficient investment.  \citet{hubert2015pipeline} and \citet{cobanli2014central} extend the geographical scope to all of Europe and Central Asia, respectively, and study how various strategic pipeline projects affect the power relation between major gas exporters, the transit countries, and the European regions importing gas. Finally,  \citet{hubert2014access}  ask how the liberalization of pipeline access within Europe affects the distribution of rents between European consumers, local incumbents, and external gas producers.

 However, the method used  in this literature to determine the value function is applicable only if all pipelines are characterized by exclusive ownership rights. With exclusive ownership, no party has access to a pipeline without the voluntary consent of the owner who is free to negotiate a compensation for his service.\footnote{Different sections of a pipeline connecting two nodes can be owned by different players (e.g.\ when pipelines cross several countries). Obviously,  all players owning a section have to grant access for the gas transport. } In the model only coalitions that include the owner of a pipeline can use it. When calculating the value of such a coalition, we can, therefore, ignore what the players outside of the coalition do. In the language of game theory, such a setting is called ``free from externalities'' and can be represented by the so called characteristic function form (CFF), assigning one real number to each possible coalition of players.

Exclusive ownership is not the only relevant institutional setting. For example, many pipelines within the European Union are subject to regulated  third party access (TPA). Since the early 1990s, the European Commission and the European Council have adopted a number of increasingly assertive directives and regulations to develop the common market for gas by ensuring fair TPA access to the transportation system within the Union --- see \citet{eu1991directive,eu1998directive,eu2003directive,EU2005-Regulation,EU2009reg}. While much of the details of access regulation have been left to the member countries, some have established a system of transport fees overseen by a regulatory authority. Under such a regime, the owner of a pipeline no longer enjoys exclusive ownership. Instead, he has to grant access, provided he is compensated according to the regulated tariff. TPA has the potential to create externalities between coalitions in two ways:
\begin{description}
\item[Resource allocation.] Suppose that the owner of a TPA-regulated pipeline is a member of a particular coalition that wants to make use of the pipeline. Under TPA, outsiders may also be entitled to use this pipeline. If the capacity of the pipeline cannot accommodate all requests, then the availability for the coalition under consideration, hence, the coalition's value, depends on the pattern of cooperation among outside players and the rules for solving conflicting claims under TPA.
 \item[Rent transfer.] Since all coalitions maximize the economic surplus, the economic costs of producing and transporting gas (e.g.\ cost of pressurizing, maintenance, etc.) are accounted for. Regulated transport fees, however, may stipulate a compensation above and beyond economic cost, for example, in order to allow the pipeline owner to recover sunk investment cost. With such a (quasi) rent transfer, the value of a coalition again depends on whether outsiders want to use resources owned by its members.\footnote{\citet{hubert2015pipeline} and \citet{hubert2014access} consider TPA for pipelines within EU, but they ignore possible capacity conflicts and assume that access fees only cover the cost of the transport of natural gas so that no player would receive a transport profit. Such a regulation would induce efficient pipeline usage and is certainly an interesting benchmark, but it is also a kind of worst-case scenario for the owners of the pipelines and politically not very likely. }
\end{description}
 In the presence of externalities that naturally arise with TPA pipelines, the surplus of a coalition  depends not only on its members, but also on the patterns of cooperation between the outside players, the so-called coalition structure or partition of players.  As a result, the gains from cooperation have to be described in partition function form (PFF) \citet{Thrall:1963}. Unfortunately, there exists neither a commonly agreed method to calculate this PFF nor to solve PFF games in general.

In this paper we use a simple example to propose a method applicable for gas networks. The iterative flow mechanism traces gas flows and transport fees for the computation of the PFF. In a second step we derive the minimal claim function based on the recursive core to obtain the corresponding CFF game. Finally, the game is solved with the Shapley Value as a measure of bargaining power. Using a simple linear framework, we show that the explicit consideration of transport fees and transport profits may significantly affect the results.

The structure of the paper is as follows.
In section \ref{MA} we summarize the model principles. In section \ref{cf} we show how the defined framework can be used for characteristic function form games. Section \ref{pf} describes the main contribution, namely, we show how the defined framework can be used for partition function form games. Section \ref{Results} demonstrates the potential of the proposed approach on a simple example. Finally, section \ref{conclusions} concludes.

\section{Model}\label{MM}

In this section we describe the principles of the model which is a generalisation of the setup introduced by \citet{Sziklai2018}.  First, we introduce the key assumptions, then, we summarize the model parameters, and finally, we describe the details of calculation.

\subsection{General Approach}\label{MA}

We consider a gas network with $l$ nodes and $m$ pipelines connecting the nodes.
Sources of natural gas (the sites of gas production) and gas consumers may be located at different nodes. We assume  $n$ players, which we associate with nodes, though, any player may hold multiple nodes ($n\leq l$).  In line with most of the literature mentioned in the introduction, one might consider the players to be countries or regions and pipelines to be long distance trunk lines. We distinguish between exclusive pipelines and TPA pipelines. A coalition may use an exclusive pipeline only if the players associated with both nodes are members of the coalition. A TPA pipeline can be used by all players, provided they pay a regulated transport fee to its owners.

 We assume inelastic demand: A fixed level of gas consumption is given in the different nodes. Any node with nonzero consumption has access to an alternative local "backstop technology" describing the cost of substituting natural gas usage. These option may represent access to LNG, alternative fuels, or even the cost of curtailing gas consumption.  Regarding these alternative sources, we assume that they cannot be traded. We assume constant unit cost for gas production, backstop technologies, and gas transport. The unit cost may differ for the various nodes and  pipelines. We also assume regulated transport fees per unit to be constant and not lower than the transport cost on the respective pipeline. We refer to the difference between the regulated fee and the transportation cost as transport profit. With these assumptions, the optimal flow problem turns into a linear cost minimization problem.

  The proposed model tracks the transport fee payments on the level of individual flows. In other words, in our model we do not only describe the 'flows of a coalition,' but we distinguish between flows corresponding to different players in the coalition. This implies that for a given coalition, we have to determine the flows of the members one by one.

  On the one hand, this will be required since we assume that the transfer fees are paid for nominal transfers, not for net transfers. The following simple example shows that, in this case, it is necessary to keep track of the individual flows. Let us assume a chain of pipelines as $A-B-C-D-E$, where $B$ and $D$ are sources, and furthermore, the middle of the pipeline, which belongs to $C$, is TPA. If in the coalition $\{A,B,D,E\}$ $A$ buys from $D$ and $E$ buys from $B$, then the nominal flows through $C$ will be higher than the net flow since the flows $D \rightarrow A$ and $B \rightarrow E$ are counter-directed in contrast to the scenario where $A$ buys from $B$ and $E$ buys from $D$. On the other hand, this way we are able to track the payments also on the level of players, not only on the level of coalitions, which may be useful in the analysis and contextualization of the results. For this, however, we need a rule for the sequential determination of the flows of the players inside a coalition.

  We follow \citet{Sziklai2018} in using an iterative flow allocation algorithm to address this problem. First, the flows of the player with the largest demand are optimized (more precisely, the node with the largest demand belonging to the player with the largest demand since multiple nodes are possible), then, the second, and so on. This potentially means that players with the smallest demand sometimes have to use more expensive sources and routes.  We conjecture that gas providers tend to prioritize large customers over small costumers, so this simplifying assumption is acceptable to keep the model complexity at an intermediate level. Note that if the optimization is performed in one step, then the individual flows may not be identified, hence the iterative approach.  Let us note here that players in a coalition pay transport fees to each other for using  pipelines.
This way transport profits also arise inside a coalition. However, these transfers within a coalition net out when calculating the coalition value.

\subsection{Formal Set-Up}
In the following we use the lower index $i$ for the indexing of players, the index $j$ for indexing nodes, and the index $k$ for indexing edges (pipelines).
The pipeline network structure is described by a directed graph and its induced node-branch incidence matrix $A$. $A\in \mathbb{R}^{l\times m}$, where $l$ is the number of nodes and $m$ is the number of edges.
Unlike  \citet{roson2015bargaining} or \citet{Sziklai2018}, where players are strictly corresponding to nodes, we do not assume bijective node-player mapping. The matrix $\Lambda \in \mathbb{R}^{l\times n}$ defines the node-player relations. $\Lambda(j,i)=1$ if and only if the node $j$ belongs to player $i$. As each node may belong to only one player, the sum of each rows in $\Lambda$ is 1.  

\subsubsection{Edge properties}

We denote the transport capacity of edge $k$ by $\bar{q}_k$. $\bar{q} \in \mathbb{R}^{m}$ is the vector of transport capacities. Transport capacities limit the maximal flow values on edges. The matrix $T\in \mathbb{R}^{n \times m}$ represents the transport costs. $T(i,k)$ denotes the cost imposed on player $i$ by the transport of one unit of gas on pipeline $k$. Similarly to the transport costs, $F\in \mathbb{R}^{n \times m}$ defines the transport fees. $F(i,k)$ is the fee that player $i$ is entitled to receive from other players for  the transport of one unit of gas on pipeline $k$.  In the following examples it is assumed that the owner(s) of the pipeline is (are) determined by the two endpoint nodes of the pipeline.
We assume that $T(i,k)\leq F(i,k)~~\forall (i,k)$. In the special case when $T(i,k) = F(i,k)$, line $k$ does not yield a profit for player $i$. The structure of $T$ and $F$ (e.g.\ the position of nonzero elements in the matrices) reflects the topographical position of pipelines and players.

\subsubsection{Node properties}

Nodal consumption is denoted by the vector $d \in \mathbb{R}^{l}$. We assume $p$ different sources for gas and define the matrix $S \in \mathbb{R}^{l\times p}$, which maps the sources to the nodes. Hence, $S(j,r)$ defines the (nonzero) unit cost of gas production from the $r$th source, found in node $j$.  $\bar{L}\in \mathbb{R}^{p}$ defines the maximal production capacity of the different sources.

\subsection{Determining the flows and the value of a coalition}
\label{cf}
In this section we determine the value of a coalition in itself (not embedded in any partition), serving as the basis for a characteristic function form (CFF) cooperative game.
Let $N$ denote the set of players corresponding to the $n$ individual players. Subsets of $N$ are called coalitions, and when treated as a coalition, $N$ is called the grand coalition.

\begin{figure}[h!]
  \centering
  \includegraphics[width=10cm]{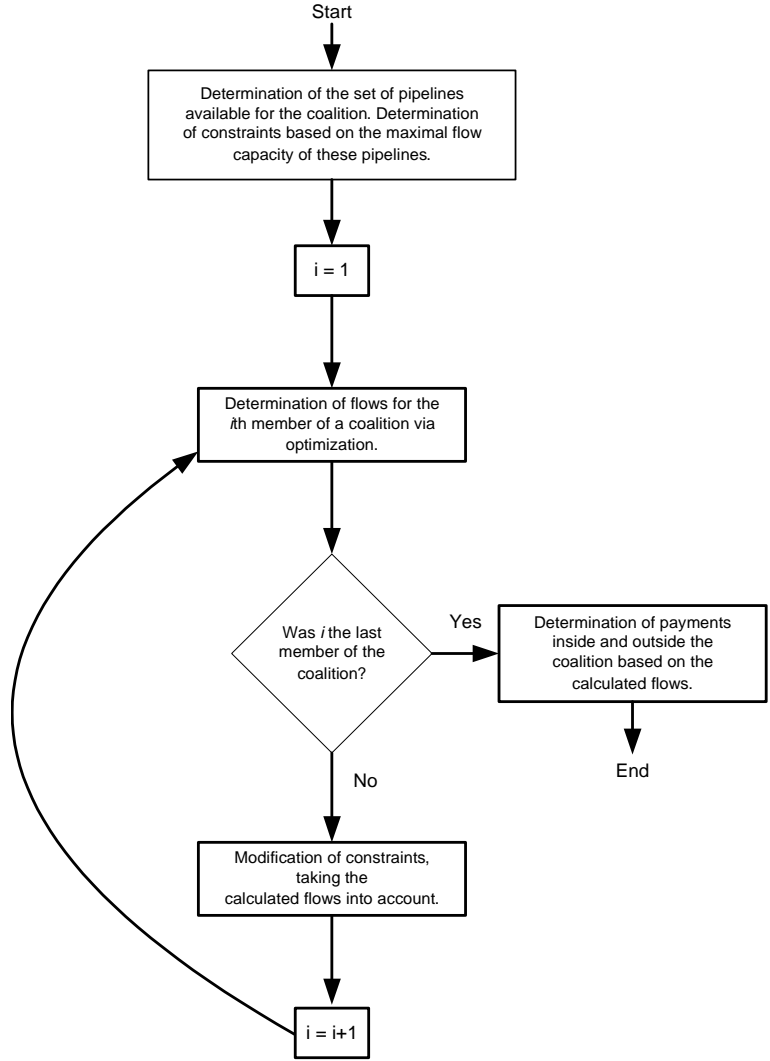}\\
  \caption{The flow chart of the calculation of gas flows for a given coalition.
  We assume that the members of the coalition are sorted in decreasing order with respect to their demands. In particular, the first member of the coalition has the maximal total demand of gas among the members.
  }
  \label{flowchart}
\end{figure}

As previously discussed in \ref{MA}, for any coalition the players' flows are determined in an iterative way, progressing from the player with the highest total consumption to other players. The process of the iterative calculation is depicted in Fig. \ref{flowchart}.

As a first step we determine which resources are accessible. This step effectively defines a sub-network. Any player in any coalition may use only gas production of nodes that belong to players in the coalition. Any coalition may use those pipelines whose both endpoints belong to players in the coalition. In addition, it may use all pipelines under TPA regulation. In the latter case the corresponding transfer fees have to be paid for all corresponding players (whose row holds a nonzero element in the matrix $F$ in the column of the pipeline).

\subsubsection{Determination of flows for the $i$th member of a coalition}

The transfer capacities and gas sources available for the $i$th member of the coalition are determined by available pipelines, total transfer capacities, and the flows of the previous members of the coalition. Under these constraints, the $i$th player satisfies his consumption, minimizing the sum of the production cost, own transport costs, transfer fees paid to other players, and (possibly) the cost of using backstop technology. In the following we will formalize the above elements.

\paragraph*{Variables and Constraints}
Under the above considerations and assuming a given coalition, every player of the coalition must solve a linear programming (LP) problem to determine its flows in the network and minimize its costs.

Variables of the LP optimization problem are as follows:   let $f_k^+$ ($k\in \{1,...,m\}$) denote the flow in the positive direction on the $k$th edge. Let $f^+ \in \mathbb{R}^m$ be the vector of these flows and similarly $f^- \in \mathbb{R}^m$ be the vector of flows opposite to the edge directions.  $L\in \mathbb{R}^{p}$ denotes the gas production at the different sources, which is constrained by the respective production capacities.

\begin{small}
\begin{equation}\label{inlet_limit}
L \leq \bar{L}
\end{equation}
\end{small}

\paragraph*{Equality type constraints}
Equality constraints may be classified into two categories: fulfillment of coalitional consumption and constraints originating from the exclusion of non-member sources.

The fulfillment of coalitional consumption is formally defined as follows.
Let us define the matrix $S^{>0}$ with the same size as $S$, holding ones on the positions where $S(i,j)>0$. Let $d_{C(i)}$ be
  a column vector of the same size as $d$ but holding nonzero values only at the positions that correspond to node(s) of the $i$th member of the coalition. At these positions the value of $d_{C(i)}$ is the same as the value of $d$.
  In this case, the equation
\begin{equation}\label{demand_fulfill}
[A~~-A~~ S^{>0}]  \left(
    \begin{array}{c}
f^+ \\
f^- \\
L
    \end{array}
  \right)=d_{C(i)}
\end{equation}
represents the nodal conservation equations and the fulfillment of the demands corresponding to the $i$th member of the coalition ($C(i)$).

Exclusion of non-member sources means that in the case of the $i$th player, further equality type constraints describe that only those sources may have nonzero inlets that are assigned to nodes corresponding to coalition member players. These can be easily determined by $\Lambda$ as follows. Let $\Lambda_C$ denote the node-player matrix for the actual coalition, in the sense that every columns in $\Lambda_C$ corresponding to non-member players are zero vectors. Let us denote the column sum (as a row vector) of any matrix $M$ by $\sum M$. In this case, the constraint

\begin{equation}
\left(1-\sum\Lambda_C^T\right)~S~L=0,
\end{equation}
where the operation $\left(1-\sum\Lambda_C^T\right)$ is performed element-wise and describes that non-coalitional production is zero.

\paragraph*{Inequality type constraints} First, we have non-negativity constraints:

  \begin{equation} \left(
    \begin{array}{c}
f^+ \\
f^- \\
L
    \end{array}
  \right) \geq 0 .\end{equation}

Then, we have inequalities describing the limited transport capacity of pipelines and production limits.
  \begin{equation}
  \left(
    \begin{array}{cc}
      I_{2m\times 2m} & 0_{2m \times p} \\
      0_{p \times 2m} & I_{p \times p} \\
    \end{array}
      \right)
    \left(
    \begin{array}{c}
f^+ \\
f^- \\
L
    \end{array}
  \right)
    \leq
    \left(
    \begin{array}{c}
    \bar{q}_{mod}\\
    \bar{L}_{mod}
    \end{array}
    \right),
  \end{equation}
  where $\bar{q}_{mod}$ is a column vector of length $2m$, which is derived from the original $\bar{q}$ vector and from the already determined flows, corresponding to previous players. Similarly, $\bar{L}_{mod}$ is a vector of length $p$, in which the gas production for previous players are subtracted from the production capacity, and $I$ is the identity matrix.
  In the first step in the case of the first player,
  $$
  \bar{q}_{mod}=\left(
                  \begin{array}{c}
                    \bar{q} \\
                    \bar{q} \\
                  \end{array}
                \right)
    ~~~
    \bar{L}_{mod}=\bar{L}.
  $$

\paragraph*{The objective function}

Let us denote the $i$th row of matrix $M$ by $M_{i,\centerdot}$ and the $i$th column of matrix $M$ by $M_{\centerdot,i}$. Let us furthermore define the matrix $F^{i0}$ for the $i$th player as a matrix identical to $F$ aside from the $i$th row, which is zero.
The construction of $F^{i0}$ refers to the consideration that no player pays transport fees to  itself. With the above notations the objective function of the actual player can be formulated as

\begin{equation}\label{celfv}
  \min_{f^+,f^-,L}~~~ [T_{i,.}+\sum F^{i0}~~T_{i,.}+\sum F^{i0}~~\sum S] \left(
    \begin{array}{c}
f^+ \\
f^- \\
L
    \end{array}
  \right)
\end{equation}

considering the constraints detailed above.
We denote the cost of player $i$---the minimum of the optimization problem---by $\varphi(C(i))$. This way the cost of any player is the sum of the transport cost from using its own pipelines, the transport fees paid to other players for using his pipelines (which is at least equal to the operational costs of the pipelines), and the cost of gas production.

The total cost of the coalition $C$ is defined as
\begin{equation}\label{cost}
\varphi(C)=\sum_{i \in C}\varphi(C(i))~.
\end{equation}

\subsubsection{Value of a coalition}
\label{section_cf}
The value of any non-singleton coalition $C$ (in itself, not embedded in any partition) is defined as the surplus induced by the cooperation, using the sum of singleton costs as reference:
\begin{equation}\label{char_fnc}
v(C)=\sum_{i \in C}\varphi(C_i)-\varphi(C)+\pi^i(C),
\end{equation}
where $C_i$ is the singleton coalition containing only player $i$. $\pi^i(C)$ denotes the total internal transport profit of coalition $C$, the difference of the total amount of transport fees the players of the coalition get from each other and the total amount of transport costs the players of the coalition imply to each other.
In other words, the intra-coalition transport profits are internalized by a coalition.
Since the above calculation can be carried out for every coalition, the characteristic function of the game is well defined.

Let us note that the sequence of the iterative procedure depicted in Fig. \ref{flowchart} does not affect the resulting value of a coalition. If the demands of the players are considered in a different order, it is possible that the flows corresponding to individual players will be different but the final picture will be the same (the most economical for the coalition). In other words, players evaluated first take the less expensive routes and sources and players evaluated in the following may take only the more expensive routes and alternatives. However, as we will see later in the case of the PFF approach described in subsection \ref{pf}, in the case of partitions, where multiple coalitions are active at the same time, the evaluation sequence of the iterative algorithm may affect the results.

In the following we demonstrate the introduced concepts on a simple example.

\subsubsection{Example 1: Transfer Profits in the Absence of TPA}
\label{Example 1}

\begin{figure}[h!]
  \centering
  \includegraphics[width=7cm]{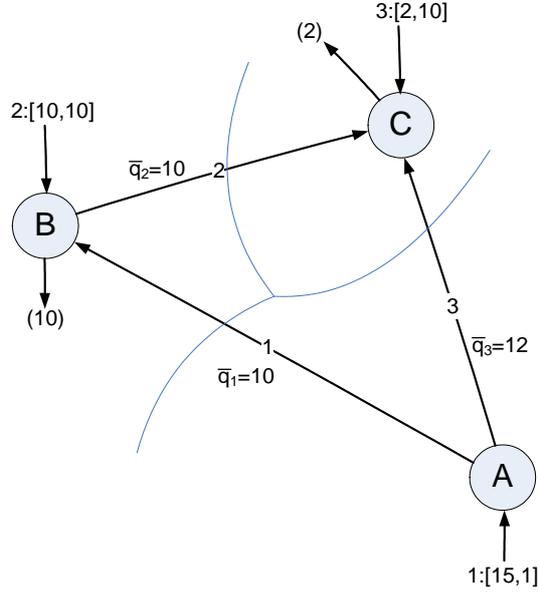}\\
  \caption{Network 1. }\label{net_example_3p_1}
\end{figure}

Let us consider the network depicted in Fig. \ref{net_example_3p_1}.

The letters in circles indicate the players (countries). Nodes with arcs leaving the node have a nonzero consumption, denoted by numbers at the end of such arcs in parentheses\footnote{Although the proposed approach is capable of handling cases where one player has multiple nodes, in this example we have a bijective node-player mapping to keep the calculations as simple as possible, while demonstrating the externalities arising from transport profits.}.
Gas production is indicated with arcs entering the node in question. The sources are numbered. After each source index, the production capacity (maximum inlet) and the cost of production is indicated in square brackets. In this example only source 1 corresponding to node $A$ is considered a 'real' source; the others correspond to backstop technologies. The $\bar{q_i}$ values indicate the maximal transport capacity of pipelines.

We have to define the matrices discussed in Section \ref{MA}. These matrices can be be found in Appendix A.
In the same Appendix in equations (\ref{T}) and (\ref{F}), the reader can find matrices $T$ and $F$ describing the transport costs and fees.

The reference costs for the players (the cost of singleton coalitions) are as follows:
\begin{equation}
\varphi(\{A\})=0~~~~\varphi(\{B\})=100~~~~\varphi(\{C\})=20.
\end{equation}
These reference costs correspond to the scenarios when no gas transport takes place, and players B and C cover their demand from their own (expensive) backstop technology.

First, we use this example to illustrate that transport fees being larger than transport costs (positive transport profits) do not matter in the absence of TPA because they net out at the coalition level. Let us consider coalition $\{A,B \}$ as an example. Player B imports 10 units of gas from player A on the direct link (edge 1). Using the $T$ and $F$ matrices in Appendix A,  we see that in this case the total expense of transfer is 80 (40 own cost and 40 paid as transfer fee to player A). As we can see in the first columns of the matrices $F$ and $T$, neither player A nor B gets any transfer profit from this transfer, since the corresponding elements are equal in the two matrices. As the source cost is 10, this means 10 units of savings compared to the reference cost of 100.

In other words, equation \ref{char_fnc} becomes
\begin{align}
v(\{A,B\})&=\varphi(\{A\}) + \varphi(\{B\}) -\varphi(\{A,B\})+\pi^i(C) \nonumber \\
\nonumber \\
& = 0 + 100 - 90 + 0=10.
\end{align}

Now we introduce positive transport profits by raising the fee above cost. Formally, we change the matrix $F$ to $F'$ as
\begin{equation}\label{F_v}
    F'=\left(
      \begin{array}{ccc}
    5 &         0   & 0.7 \\
    4  &  0.5       &  0 \\
         0  &  1 &    1
      \end{array}
    \right).
\end{equation}
The cost of coalition $\{A,B\}$ would change, as in this case player B pays 50 units to player A for the transport of 10 units. But in this case, a transport profit of 10 units is also generated, since as described by the matrix $T$, the transport implies still only a cost of 40 units to player A. As player A is in the coalition, both effects net out and equation \ref{char_fnc} becomes
\begin{align}
v(\{A,B\})&=\varphi(\{A\}) + \varphi(\{B\}) -\varphi(\{A,B\})+\pi^i(C) \nonumber \\
\nonumber \\
& = 0 + 100 - 100 +10 =10,
\end{align}
giving the coalition with the same payoff. Transport profits can play a role only in the case of some pipelines being subject to TPA regulation.


\subsection{Determining the flows and the values of a partition}
\label{pf}
In this subsection we extend the analysis to situations in which TPA regulation creates externalities, giving rise to a partition function form (PFF) game.

 A partition of $S \subseteq N$ is a set of mutually disjoint nonempty coalitions whose union is $S$ ($N$ is the player set).  Let $\Pi (S)$ denote the set of partitions of $S\subseteq N$.
An embedded coalition is a pair $(C,\PP)$, where $C\in\PP \in \Pi(N)$. The set of embedded coalitions is denoted by $\EE$. A game in partition function form (\citet{Thrall:1963}) is a pair $(K,V)$, where $V:\EE \rightarrow \mathds{R}$ is the partition function, which assigns a real payoff to each embedded coalition.

This enables us to describe the positive externalites of transport profits formally. For example, according to the approach detailed in the previous subsection, if player $X$ is not cooperating with anyone and no TPA pipelines are present, then basically its value is 0. Embedded in a partition, however, where $Y$ and $Z$ are cooperating and using pipelines crossing the territory of $X$ (thus TPA is present), $X$ may profit from the transport depending on the respective values of $T$ and $F$.

Our iterative approach is analogous to the case of simple coalitions. First, we identify the coalition with the greatest total demand in the partition and calculate its flows and value, as in section \ref{cf}. However, in this case, we calculate the externalities implied by the coalition and update the value of outside coalitions (which may be singletons) as well. Then, we calculate the flows of the next coalition, taking the flows determined in the first step into account. After the calculation for each coalition, we update the values of every coalition value. The externality implied by coalition $C_1$ to coalition $C_2$ is the sum of the externalities $C_1$ implies to the members of $C_2$.

We have to note that in special cases when multiple coalitions with different gas sources are active at the the same time and the transport capacity of the commonly used pipelines is limited, it may be important which coalition we evaluate first (the one with the highest total demand or another). This phenomena is demonstrated with an example in Appendix B.

Externalities of any coalition $C$ to others may be easily determined based on the following 2 matrices:

\begin{itemize}
  \item $Q_{\PP}(i,j)$ is the transport cost imposed by player $j$ on player $i$, assuming the partition $\PP$. The $j$th column of $Q_{\PP}$ is determined as
\begin{equation}
Q_{\PP}(.,j)=[T~~T~~0_{n \times p}]x_j~~~j\in C~,
\end{equation}
where $x_j$ is the solution of the optimization problem of player $j$ in the case of the given partition, composed of $f^+$, $f^-$ and $L$ corresponding to the actual player.
  \item $R_{\PP}(i,j)$ is the transport fee paid by player $j$ to player $i$, assuming the partition $\PP$. The $j$th column of $R_{\PP}$ is determined as
  \begin{equation}
R_{\PP}(.,j)=[F^{j0}~~F^{j0}~~0_{n \times p}]x_j~~~j\in C~.
\end{equation}
\end{itemize}

If we take the difference of the two matrices above and calculate the row sums for rows corresponding to players outside of $C$, we
get the externalities implied by $C$ at the player level. The externalities caused to any coalition $C'$ may be derived by summing up the externalities of the players in $C'$.

\subsubsection{Value of a coalition embedded in a partition}
\label{section_pf}
The value of any non-singleton coalition $C$ embedded in partition $\PP$, denoted by $v_{\PP}(C)$, is defined as the surplus achieved by the cooperation, using the sum of singleton costs as reference (see equation (\ref{part_fnc})) and the externalities implied to it by the transport of other coalitions:
\begin{equation}\label{part_fnc}
v_{\PP}(C)=\sum_{i \in C}\varphi_{\SSS}(C_i)-\varphi_{\PP}(C)+\pi(C),
\end{equation}
where $C_i$ is still the singleton coalition containing only player $i$. $\varphi_{\SSS}(C_i)$ is the cost of the singleton coalition $C_i$ embedded into the partition $\SSS$, which is the partition holding only singleton coalitions. $\varphi_{\PP}(C)$ denotes the cost of coalition $C$ embedded in partition $\PP$.
Appendix B gives a simple example of TPA-induced resource allocation and shows how the cost of a coalition embedded in a partition may be affected by the partition.

Furthermore, in the case when the coalition is embedded in a partition, $\pi$ in equation (\ref{part_fnc}) differs from $\pi^i$ in equation (\ref{char_fnc}), as it describes not only internal transport profits, but also transport profits received from other coalitions. Formally, $\pi=\pi^i+\pi^e$, where the new term $\pi^e$ is the transport profit originating from external transports.

To show a simple example calculation of how externalities are generated, we turn to example 1 again, depicted in Fig. \ref{net_example_3p_1}. Let us consider the partition $\PP_1=\{ \{A,B\},\{C\}\}$  and TPA regulation for lines 2 and 3.

With the lines between nodes B-C and A-C now being accessible, the transport of coalition $\{A,B\}$ takes place via player C since this is the more economic alternative.
In this case, the matrices $Q_{\PP_1}$ and $R_{\PP_1}$ will be as follows:

\begin{equation}
Q_{\PP_1}=\left(
  \begin{array}{ccc}
     0  &   7  &  0 \\
     0  &   5  &  0 \\
     0  &   8  &   0 \\
  \end{array}
\right)~~~
R_{\PP_1}=\left(
  \begin{array}{ccc}
     0  &   7  &  0 \\
     0  &   0  &  0 \\
     0  &   20  &   0 \\
  \end{array}
\right)~~~.
\end{equation}

In this cooperation structure, only player $B$ has transports, which is why the second columns of the matrices $Q_{\PP_1}$ and $R_{\PP_1}$ are nonzero. Now, as described before, to calculate the externality from coalition $\{ A,B\}$ to $\{C\}$, we take the third row (corresponding to player $C$) of the matrix $R_{\PP_1}-Q_{\PP_1}$, which is equal to 12.
In other words, $B$ pays a transport fee of 20 units to player $C$, from which 5+3=8 units is the transport cost, and thus, it generates a transport profit of 12.


Since the above calculation can be carried out for every coalition embedded in any partition, the partition function of the game is well defined.

\subsection{Analysis of PFF games}\label{sec:PFF}

The PFF game obtained by the procedure described in the previous subsections describes the gains from cooperation when regulated TPA creates externalities. There are two approaches to complete the analysis by deriving a power index for the different players. First, one might straightforward solve the PFF game as such. While some solutions for PFF games have been proposed,  e.g.\ the extended Shapley-value, these are much more complex (see \citet{mcquillin2009extended}) than for a CFF game. Hence, as a second option, one might use additional considerations to select a unique partition for every possible coalition and take its value under this particular partition to be the value for the CFF game.

Here we use the concept of the \emph{minimal claim function} defined by \citet{habis2014cooperation} to derive the characteristic function. The basic concept of the minimal claim function is that each coalition's minimal claim is equal to the value the coalition gets when the rest of the players form a  partition that is stable in the spirit of the recursive core \citep{Koczy_recursive} \footnote{More precisely, when more stable partitions exist for the residual game, the one implying the least payoff for the coalition in question will derive the minimal claim value of the coalition. In other words, the pessimistic recursive core is used.}. An example demonstrating the calculation of the minimal claim function is described in Appendix C. We denote the minimal claim function by $v_{MC}$.

As alternatives for the recursive core, the optimistic core stability concept of \citet{Shenoy1979}, the Gamma-core of \citet{Chander2007}, and the pessimistic $\alpha$-core of \citet{Aumann1960} may be used as well. While in the former setup a coalition deviates if it has any chance of higher payoff, in the latter a coalition deviates only if it gets a higher payoff irrespective of the cooperation structure of the other players.
For a comparative discussion of the different approaches, see \citet{Koczy2018}.


\subsection{Shapley value}\label{sec:Shapley}

Finally, we have to decide how to solve the CFF game obtained by applying the minimal claim function to the underlying PFF game. Arguably, the Shapley value \citep{Shapley1953} is the best-known single-value solution concept in cooperative game theory (\citet{roth1988shapley}). Its popularity comes partly from the fact that there is an explicit formula to compute it, partly from its very convincing axiomatic characterizations. \cite{Young1985} shows that the Shapley value is the only solution which satisfies Pareto-optimality, Symmetry, and Strong monotonicity. This latter axiom entails that if a player's marginal contributions in one game are consistently as large as the same marginal contributions in another game on the same player set, then the player's payoff in the first game should be at least as large as in the second game. In other words, the Shapley value rewards productivity. This property makes the Shapley value an excellent candidate for a power measure.

In contrast, the nucleolus \citep{Schmeidler1969}, another popular cooperative solution concept, received less attention as a power measure. The nucleolus is obtained via a lexicographic optimization, during which the profits of the poorest coalitions are improved until a unique solution is obtained. In this sense, the nucleolus implements some kind of social justice, which has little to do with the bargaining positions of the players. Although \citet{Montero2005} argues that the nucleolus may outperform the Shapley value as a power index for simple voting games, some empirical evidence shows that concerning gas markets the Shapley value has the better explanatory power. \citet{hubert2011investment} argue that terms of contracts between Russia and the transit countries Ukraine and Belarus around 2002-2003 can be explained using the Shapley value but are at odds with predictions by the nucleolus and the core\footnote{The core is a set valued solution concept which contains the stable allocations. The nucleolus is always a core member whenever the core is non-empty, however, the Shapley value may lie outside the core.}. In their analysis of three major pipeline projects, \citet{hubert2015pipeline} can explain the successes and the failures, assuming that the players use the Shapley value to assess their power in the network, while the nucleolus yields predictions which contradict observable investment behavior.

\section{Results and discussion}
\label{Results}

In this section we first complete all steps of the analysis for the simple network described in \ref{Example 1}. In doing so we demonstrate that explicitly considering transport profits as externalities may give significantly different results compared to conventional approaches of cooperative game theory. Second, we compare the different results for a more complex and more realistic example describing the transport of natural gas from Russia to Central Europe via Ukraine.

\subsection{Example 1}

In the following we compare the traditional CFF approach (see e.g.\ \citet{hubert2011investment},  \citet{hubert2008dynamic},  \citet{hubert2014access}, \citet{hubert2015pipeline},  \citet{cobanli2014central}, \citet{roson2015bargaining}) and the PFF approach proposed in this article with and without TPA. In the case of the PFF approach, after the calculation of the partition function we derive the recursive-core-based minimal claim function, with which we calculate the Shapley value and compare this with the Shapley value derived from the simple CFF approach. We use the 3-node network introduced in subsection \ref{Example 1}.

\subsubsection{No third party access (TPA)}

 The value of $\{A,B \}$ is discussed in subsection \ref{Example 1}.
 The value of coalition $\{A,C \}$ may be derived similarly.
In the case of the grand coalition, all transports take place via the route through player C. In this case, player B pays 10+10 units of transport fee to player C for using lines 2 and 3, respectively, implying a transport profit of 5+7=12 to player C. Further costs for player A include the 7 units of transport fee paid to player A, the transport cost of 5 corresponding to line 2, and the 10 units of gas production cost. This means a saving of 58 units for player B. Player C pays 1.4 units of transport fee to player A along with 0.6 transport cost and 2 units of production cost, resulting in a saving of 16. As in this case the transport profit of player C is internalized, the value of the coalition is equal to 58+16+12=86.
Thus, the CFF scenario results in a simple characteristic function described in Table \ref{T1} (singletons have zero value in this case).

\begin{table}[h!]
  \begin{tabular}{c|c|c|c|c}
    coalition & \{A,B\} & \{A,C\} & \{B,C\} & \{A,B,C\} \\ \hline
    value & 10 & 16 & 0 & 86
  \end{tabular}
  \caption{The characteristic function of the network depicted in Fig. \ref{net_example_3p_1} if no TPA is allowed. \label{T1}}
\end{table}

If we use the PFF approach, then using similar considerations as above, we get the partition function described in Table \ref{T2}.

\begin{table}[h!]
  \begin{tabular}{c|c|c|c}
    $\PP$ &  $v(C)$ &     $\PP$ &  $v(C)$  \\ \hline
	\{A,B,C\} & 86 & 	\{A,B\} + \{C\} & 10 , {0}\\  \hline
	\{A,C\} + \{B\} & 16 , 0 & 	\{A\} + \{B,C\} & 0 , {0}\\ \hline
	\{A\} + \{B\} + \{C\} & 0 , 0 , 0 &
  \end{tabular}
    \caption{The partition function of the network depicted in Fig. \ref{net_example_3p_1} if no TPA is allowed. \label{T2}
    $\PP$ denotes the actual partition, while $v(C)$ is the vector of values of coalitions embedded in the particular partition.}
  \end{table}

When using the recursive-core-based approach, we obtain a the minimal claim function which is identical to the characteristic function described in Table \ref{T1}. This illustrates that without TPA, the CFF and PFF approaches coincide since transport profit is only produced inside of coalitions in this case. The Shapley values derived in this case are [33, 25, 28] for players A, B, and C, respectively, which also coincide with the extended Shapley value for the partition function.

 \subsubsection{TPA of lines 2 and 3}

The CFF approach results in the characteristic function described in Table \ref{T3} and implies the Shapley values [41, 33, 12]. As we can see, the TPA of lines 2 and 3 significantly decreased the bargaining power of player C, which is not surprising.

\begin{table}[h!]
  \begin{tabular}{c|c|c|c|c}
    coalition & \{A,B\} & \{A,C\} & \{B,C\} & \{A,B,C\} \\ \hline
    value & 58 & 16 & 0 & 86
  \end{tabular}
  \caption{The characteristic function of the network depicted in Fig. \ref{net_example_3p_1} if TPA of lines 2 and 3 is assumed. \label{T3}}
\end{table}

The partition function, on the other hand, is summarized in Table \ref{T4}. If we derive the extended Shapley value (\citet{mcquillin2009extended}) for the partition function, we get [39, 31, 16].

\begin{table}[h!]
  \begin{tabular}{c|c|c|c}
    $\PP$ &  $v(C)$ &     $\PP$ &  $v(C)$  \\ \hline
	\{A,B,C\} & 86 & 	\{A,B\} + \{C\} & 58 , {12}\\  \hline
	\{A,C\} + \{B\} & 16 , 0 & 	\{A\} + \{B,C\} & 0 , {0}\\ \hline
	\{A\} + \{B\} + \{C\} & 0 , 0 , 0 &
  \end{tabular}
    \caption{The partition function of the network depicted in Fig. \ref{net_example_3p_1} if TPA of lines 2 and 3 is assumed. \label{T4}}
  \end{table}

Regarding the recursive-core-based minimal claim function approach, the minimal claim function derived from the partition function is described in Table \ref{T5}.
The most important point is that in this case, the value of the singleton coalition $C$ is nonzero (because of the transport profit corresponding to the transport between A and B).

\begin{table}[h!]
  \begin{tabular}{c|c|c|c|c|c|c|c}
    coalition &\{A\}& \{B\} & \{C\} & \{A,B \} & \{A,C \} & \{B,C \} & \{A,B,C \} \\ \hline
    MC value & 0 & 0 & 12 & 58 & 16 & 0 & 86
  \end{tabular}
  \caption{The minimal claim function of the network depicted in Fig. \ref{net_example_3p_1} if TPA of lines 2 and 3 is assumed. \label{T5}}
\end{table}

If we derive the (conventional) Shapley value from the minimal claim function, then we get [39, 31, 16] in this case, which coincides with the extended Shapley value.
As we can see, the PFF approach, irrespective of which Shapley value calculation we use, reflects how player 3 benefits from the transport profit, as his Shapley value is increased by 33\%.

\subsection{Example 2}
\begin{figure}[h!]
  \centering
  \includegraphics[width=9cm]{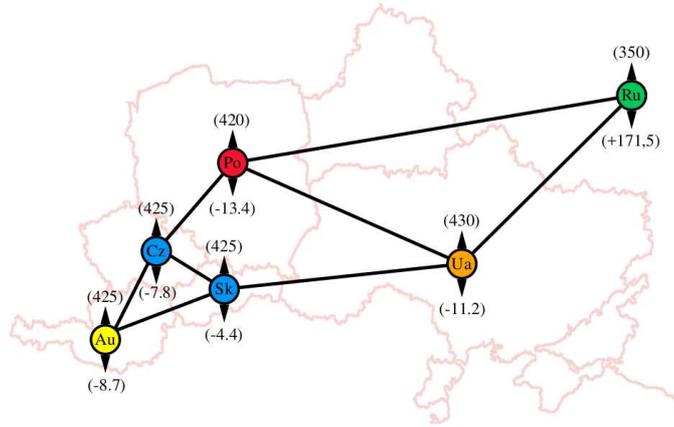}\\
  \caption{Network of example 2.}\label{rl_example2}
\end{figure}

In this example we study the path of the natural gas from Russia to Central Europe.
We define the following players to study this scenario: Poland (Po), Austria (Au), Czech Republic and Slovakia (Cz-Sk), Ukraine (Ua), and Russia (Ru). The 3rd player representing the Czech Republic and Slovakia together holds two nodes, while the other players hold one node each. The network is depicted in Fig~\ref{rl_example2}. Numbers on the top of the nodes represent the cost of production/alternative source, while the numbers on the bottom show the amount of gas produced/demanded. Further numerical parameters may be found in Appendix D.

The pipeline between Russia and Poland crosses the territory of Belarus.  However, this segment of the Yamal-Europe pipeline  is owned by GAZPROM. To keep the example simple we do not take Belarus into account as a separate player. The countries in the example consume mainly (although not exclusively) Russian gas, thus, we omitted other suppliers as strategic players. We assume that if Russian gas is not available, then the countries use alternative sources, such as LNG, which cost somewhat more. The parameters of the network, such as nodal demand, pipeline capacities, etc.\ were compiled from \citep{Sziklai2018}. The production cost and transport fees in the example were determined based on \cite{weiner2016central} as follows:

\begin{itemize}
\item Russian gas production cost was determined by taking the German import price and subtracting the transportation cost via the Nord Stream pipeline.
\item To determine the transport fees of Ukraine we took the gas import price of  Hungary and subtracted the Russian gas production price and related transport costs.
\item The base price of the alternative source was set as 20\% more expensive than the Russian gas. This was further adjusted with the network position, that is, the price is increasing from West to East and from North to South.
\end{itemize}

In addition to the above, we made the following assumptions:

\begin{itemize}
\item All lines in the network are taken into account as TPA.
\item We assume that alternative sources may always be transported to the node in question directly. 
\item We assume transport margin only on the Ukraine-Slovakia pipeline.
\end{itemize}

\begin{table}[h!]
\begin{center}
\begin{tabular}{|c|c|c|c|c|c|}
  \hline
   & Po & Au & Cz-Sk & Ua & Ru \\ \hline
  Shapley values based on CFF & 400.4   & 182.7  &  247.9  &  663.0    &1263.7 \\
  Shapley values based on PFF & 400.4  &  134.3  &  179.0  &  894.7 &   1149.2\\
  \hline
\end{tabular}
\caption{Shapley values of example 2 in the case of the traditional characteristic function form approach (CFF) and the proposed partition function form approach (PFF)}\label{table_ex_2}
\end{center}
\end{table}

Table \ref{table_ex_2} displays the Shapley values derived from the CFF and PFF approaches. We can see that applying the PFF approach and explicitly considering the transport profits results in a different characterization of bargaining power. In particular, the Shapley value of Ukraine is nearly 35\% greater in the PFF game than in the CFF game and Ukraine's position relative to Russia is significantly improved.

We have discussed before that the sequence of the iterative process may affect the results in the case of the PFF approach. As only one real source (Russia) is present in the network, scenarios like in Appendix B do not arise in this case, and thus, the result is independent of the sequence of the iterative algorithm.

\section{Conclusions}
\label{conclusions}

In this paper we define a framework for the cooperative game-theoretic modeling of gas networks and markets, able to explicitly consider externalities that may arise due to regulated TPA. These externalities have been largely ignored in the previous cooperative literature on gas networks. Regulated TPA creates two challenges for the traditional approach: First, it may lead to incompatible claims on scarce pipeline capacity and second, it may result in transfer payments for transport services across coalitions.

We propose a solution for the second challenge by  determining the flows of a coalition in an iterative way that tracks transport fees in a two-party, contract-driven manner regarding the source and the destination of gas. This feature is essential for the calculation of transport fees and resulting transport profits since members of a coalition may imply counter-directed flows on a TPA pipeline generating transport profit for an external player. In addition, this methodology allows us to calculate the individual costs and profits of players in a given coalition, which is valuable additional information not prevalent in existing cooperative game theoretic models. Using this iterative flow mechanism, we are able to compute the partition function. In this paper we propose to solve the resulting PFF game in two steps: first, by using the minimal claim function to obtain a CFF-game and then, solving this game with the traditional Shapley Value. Two examples illustrate the methodology and demonstrate that accounting for TPA-related externalities may substantially alter the power structure.

On the other hand, as demonstrated in the case of the example described in Appendix B, the evaluation order of the coalitions may be important if TPA leads to conflicting claims on transport capacities. For these cases, it would be nice to have a more explicit mechanism for the allocation of pipeline capacities. This issue, however, is left for further research.

\section{Acknowledgements}

The authors thank the funding of the National Research, Development and Innovation Office (K109354, K119930, K128573 and PD 123900), P\'{a}zm\'{a}ny P\'{e}ter Catholic University (KAP18-1.1-ITK), and the Hungarian Academy of Sciences under its Momentum Programme (LD-004/2010). This research was supported by the Higher Education Institutional Excellence Program of the Ministry of Human Capacities in the framework of the 'Financial and Retail Services' research project ( 20764-3/2018/FEKUTSRTAT) at the Corvinus University of Budapest.

\section*{Appendix A}
Parameters of Example 1 depicted in Fig. \ref{net_example_3p_1}:

\begin{small}
\begin{equation}\label{A_lambda_barq}
A=\left(
    \begin{array}{ccc}
 -1  &   0  &  -1 \\
     1  &  -1   &  0 \\
     0   &  1  &   1
    \end{array}
  \right)
\Lambda=\left(
  \begin{array}{ccc}
1 &  0 &  0  \\
0 &  1 &  0 \\
0 &  0 &  1
  \end{array}
\right)
\bar{q}=\left(
          \begin{array}{c}
10  \\
10  \\
12
          \end{array}
        \right)
\end{equation}

\begin{equation}\label{T}
  T=\left(
      \begin{array}{ccc}
    4 &         0   & 0.7 \\
    4  &  0.5       &  0 \\
         0  &  0.5 &    0.3
      \end{array}
    \right)
\end{equation}
\begin{equation}\label{F}
    F=\left(
      \begin{array}{ccc}
    4 &         0   & 0.7 \\
    4  &  0.5       &  0 \\
         0  &  1 &    1
      \end{array}
    \right)
\end{equation}
\begin{equation}\label{d}
d=\left(
    \begin{array}{c}
0  \\
10  \\
2
    \end{array}
  \right)
  S=\left(
      \begin{array}{ccc}
   1  &   0  &   0  \\
     0 &   10   &  0 \\
     0   &  0 &   10
      \end{array}
    \right)
    \bar{L}=\left(
              \begin{array}{c}
 15  \\
 10  \\
  2
                \end{array}
            \right)
\end{equation}

\end{small}

\section*{Appendix B}
\begin{figure}[h!]
  \centering
  \includegraphics[width=7cm]{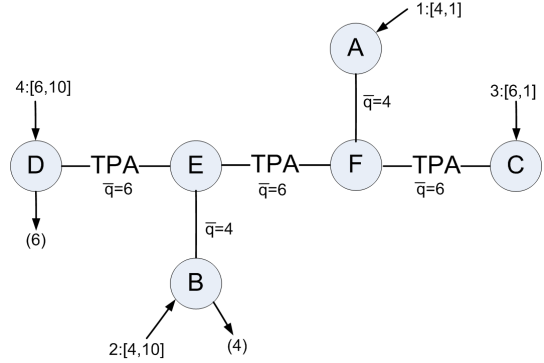}\\
  \caption{Network 2. The horizontal pipelines are TPA; E and F are dummy nodes with no consumption and sources.}\label{EE_example_2}
\end{figure}

To demonstrate how the partition in which the coalition is embedded plays a role in the determination of its cost, let us consider the example depicted in Fig. \ref{EE_example_2}. This is a simple example of the TPA-induced resource allocation discussed in section \ref{Introduction}.

Let us consider the coalition $\{A,B\}$, first embedded in the partition \\ $\{A,B\},\{C\} ,\{D\},\{E\},\{F\}$. In this case, B is able to get gas from A, so its cost is 4. Embedded in the partition
$\{A,B\},\{C,D\},\{E\},\{F\}$, the coalition with the largest total demand is $\{C,D\}$, so first the flows belonging to them are evaluated. The transport between C and D will use up all the transport capacity of the TPA pipeline between E and F, so B must use its own expensive source, so its cost will be 40.

This example also highlights how the exact sequence of the iterated algorithm described in subsection \ref{pf} may influence the values of a partition. If the flows of the coalition with the highest demand ($\{C,D\}$) are evaluated first, then the result is a flow of 6 units from $C$ to $D$. As the reference costs (assuming all singletons) for the coalitions $\{A,B\}$ and $\{C,D\}$ are 40 and 60, respectively, if we assume no transport costs and fees, then the total savings for coalitions $\{A,B\}$ and $\{C,D\}$ (and, thus, the values of the coalitions) are 0 and 54, respectively.

On the other hand, if the flows of coalition $\{A,B\}$ are evaluated first, then the resulting flows are as follows: We have a flow of 4 units from A to B and a flow of 2 units from C to D in this case, which results in savings of 36 and 18 for
$\{A,B\}$ and $\{C,D\}$, respectively.

\section*{Appendix C}
In this appendix we provide a simple example describing the calculation of the minimal claim function (in the spirit of the pessimistic recursive core).

Let us consider the partition function described in Table \ref{example_PF}.
\begin{table}[h!]
\begin{center}
\begin{tabular}{|c|c|}
  \hline
  Partition & Values  \\ \hline
  \{A\},\{B\},\{C\} & 0,3,0 \\
  \{A,B\},\{C\} & 2,1 \\
  \{A,C\},\{B\} & 2,2 \\
  \{A\},\{B,C\} & 1,2 \\
  \{A,B,C\} & 4 \\
  \hline
\end{tabular}
\end{center}
\caption{An example partition function \label{example_PF}}
\end{table}

When determining e.g. the minimal claim of the singleton coalition $\{ A\}$, we have to consider the partitions in which this coalition is embedded. This means two cases: $\{\{A\},\{B\},\{C\}\}$ and $\{\{A\},\{B,C\}\}$. We have to consider the residual game, namely the game of players $B$ and $C$. The residual players form a stable partition in the residual game if they form the partition $\{B\},\{C\}$. The cooperation $\{B,C\}$ is not stable in the residual game since the payoff of the coalition $\{B,C\}$ is only 2 in this case, and thus, player $B$ is blocking it with a payoff of 3 in the case of $\{B\},\{C\}$, resulting in the partition of $\{\{A\},\{B\},\{C\}\}$. According this we assign the value 0 to coalition $\{A\}$ as minimal claim, as this is its payoff in the partition when the players of the residual game form a stable partition. The pessimistic approach means that in the case when more stable partitions are present in the residual game, thus, more values are assigned to the coalition in question with the above approach, we choose the lowest of them.

\section*{Appendix D}
Parameters of Example 2 depicted in Fig. \ref{rl_example2}:

\begin{small}
\begin{equation}\label{A_ex2}
A=\left(
    \begin{array}{cccccccc}
     0 &  1 &  1 &  1 &  0 &  0 &  0 &  0  \\
     0 &  0 &  0 &  0 &  1 &  0 &  1 &  0  \\
     0 &  0 &  0 & -1 & -1 &  1 &  0 &  0  \\
     0 &  0 &  0 &  0 &  0 & -1 & -1 &  1  \\
     1 &  0 & -1 &  0 &  0 &  0 &  0 & -1  \\
    -1 & -1 &  0 &  0 &  0 &  0 &  0 &  0
    \end{array}
  \right)
\end{equation}
\begin{equation}\label{lambda_ex2}
\Lambda=\left(
  \begin{array}{ccccc}
    1 &  0 &  0 &  0 &  0  \\
    0 &  1 &  0 &  0 &  0  \\
    0 &  0 &  1 &  0 &  0  \\
    0 &  0 &  1 &  0 &  0  \\
    0 &  0 &  0 &  1 &  0  \\
    0 &  0 &  0 &  0 &  1
  \end{array}
\right)
\bar{q}=\left(
          \begin{array}{c}
      145  \\
     40.1  \\
        5  \\
     0.93  \\
        0  \\
    54.93  \\
       49  \\
     97.7  \\
          \end{array}
        \right)
\end{equation}

\begin{equation}\label{T_ex2}
  T=\left(
      \begin{array}{cccccccc}
    0 &      5.1225 &        5.85 &        4.65 &           0 &           0 &           0 &           0  \\
      0 &           0 &           0 &           0 &       2.175 &           0 &         0.6 &           0  \\
      0 &           0 &           0 &        4.65 &       2.175 &        4.95 &         0.6 &       8.775  \\
    1.5 &           0 &        5.85 &           0 &           0 &           0 &           0 &       8.775  \\
    1.5 &      5.1225 &           0 &           0 &           0 &           0 &           0 &           0
      \end{array}
    \right)
\end{equation}
\begin{equation}\label{F_ex2}
    F=\left(
      \begin{array}{cccccccc}
      0 &       5.1225 &         5.85 &         4.65 &            0 &            0 &            0 &            0  \\
      0 &            0 &            0 &            0 &        2.175 &            0 &          0.6 &            0  \\
      0 &            0 &            0 &         4.65 &        2.175 &         4.95 &          0.6 &        8.775  \\
    1.5 &            0 &         5.85 &            0 &            0 &            0 &            0 &           44  \\
    1.5 &       5.1225 &            0 &            0 &            0 &            0 &            0 &            0  \\
      \end{array}
    \right)
\end{equation}
\begin{equation}\label{d_ex2}
d=\left(
    \begin{array}{c}
    13.4  \\
     8.7  \\
     7.8  \\
     4.4  \\
    11.2  \\
       0
    \end{array}
  \right)
  S=\left(
      \begin{array}{cccccc}
    420 &    0 &    0 &    0 &    0 &    0  \\
      0 &  425 &    0 &    0 &    0 &    0  \\
      0 &    0 &  425 &    0 &    0 &    0  \\
      0 &    0 &    0 &  425 &    0 &    0  \\
      0 &    0 &    0 &    0 &  430 &    0  \\
      0 &    0 &    0 &    0 &    0 &  350  \\
      \end{array}
    \right)
    \bar{L}=\left(
              \begin{array}{c}
     13.4  \\
      8.7  \\
      7.8  \\
      4.4  \\
     11.2  \\
    171.5  \\
                \end{array}
            \right)
\end{equation}

\end{small}

\newpage
\bibliographystyle{apalike}
\section*{Bibliography}

 \bibliography{GT_gas_1}

\begin{thebibliography}{}

\bibitem[Abada et~al., 2013]{Abada2013}
Abada, I., Gabriel, S., Briat, V., and Massol, O. (2013).
\newblock A generalized {Nash} -- {Cournot} model for the {Northwestern}
  {European} natural gas markets with a fuel substitution demand function:
  {The} {GaMMES} model.
\newblock {\em Networks and Spatial Economics}, 13(1):1--42.

\bibitem[Aumann and Peleg, 1960]{Aumann1960}
Aumann, R.~J. and Peleg, B. (1960).
\newblock {Von Neumann-Morgenstern} solutions to cooperative games without side
  payments.
\newblock {\em Bulletin of the American Mathematical Society}, 66:173--179.

\bibitem[Boots et~al., 2004]{boots2004trading}
Boots, M.~G., Rijkers, F.~A., and Hobbs, B.~F. (2004).
\newblock {Trading in the downstream European gas market: A successive
  oligopoly approach}.
\newblock {\em Energy Journal}, 25(3):73--102.

\bibitem[Breton and Zaccour, 2001]{breton2001equilibria}
Breton, M. and Zaccour, G. (2001).
\newblock Equilibria in an asymmetric duopoly facing a security constraint.
\newblock {\em Energy Economics}, 23(4):457--475.

\bibitem[Chander, 2007]{Chander2007}
Chander, P. (2007).
\newblock The gamma-core and coalition formation.
\newblock {\em International Journal of Game Theory}, 35(4):539--556.

\bibitem[Cobanli, 2014]{cobanli2014central}
Cobanli, O. (2014).
\newblock Central {Asian} gas in {Eurasian} power game.
\newblock {\em Energy Policy}, 68:348--370.

\bibitem[De~Wolf and Smeers, 1997]{deWolf1997stochastic}
De~Wolf, D. and Smeers, Y. (1997).
\newblock A stochastic version of a {Stackelberg-Nash-Cournot} equilibrium
  model.
\newblock {\em Management Science}, 43(2):190--197.

\bibitem[{Energy Charter Secretariat}, 2007]{ECS2007-Regulation}
{Energy Charter Secretariat} (2007).
\newblock Putting a price on energy - international pricing mechanisms for oil
  and gas.
\newblock Technical report.

\bibitem[EU, 1991]{eu1991directive}
EU (1991).
\newblock {Council Directive 91/296/EEC on the transit of natural gas through
  grids}.
\newblock {\em Official Journal of the European Communities}, L147(37).

\bibitem[EU, 1998]{eu1998directive}
EU (1998).
\newblock Directive 98/30/{EC} of the {E}uropean {P}arliament and of the
  {C}ouncil of 22 {J}une 1998 concerning common rules for the internal market
  in natural gas.
\newblock {\em Official Journal of the European Communities}, L204.

\bibitem[EU, 2003]{eu2003directive}
EU (2003).
\newblock {Directive 2003/55/EC of the European Parliament and of the Council
  of 26 June 2003 concerning common rules for the internal market in natural
  gas and repealing Directive 98/30/EC}.
\newblock {\em Official Journal of the European Union}, L176.

\bibitem[EU, 2005]{EU2005-Regulation}
EU (2005).
\newblock Regulation (ec) no 1775/2005 of the european parliament and of the
  council of 28 september 2005 on conditions for access to the natural gas
  transmission networks.
\newblock {\em Official Journal of the European Union}, L289(1).

\bibitem[EU, 2009]{EU2009reg}
EU (2009).
\newblock {Regulation (EC) No 715/2009 of the European Parliament and of the
  Council on conditions for access to the natural gas transmission networks and
  repealing Regulation (EC) No 1775/2005}.
\newblock {\em Official Journal of the European Union}, L211(36).

\bibitem[Gabriel et~al., 2004]{gabriel2004nash}
Gabriel, S., Zhuang, J., and Kiet, S. (2004).
\newblock A nash-cournot model for the {North American} natural gas market.
\newblock In {\em IAEE Conference Proceedings, Zurich Switzerland}.

\bibitem[Gabriel et~al., 2005a]{gabriel2005mixed}
Gabriel, S.~A., Kiet, S., and Zhuang, J. (2005a).
\newblock A mixed complementarity-based equilibrium model of natural gas
  markets.
\newblock {\em Operations Research}, 53(5):799--818.

\bibitem[Gabriel et~al., 2005b]{gabriel2005large}
Gabriel, S.~A., Zhuang, J., and Kiet, S. (2005b).
\newblock A large-scale linear complementarity model of the {North American}
  natural gas market.
\newblock {\em Energy economics}, 27(4):639--665.

\bibitem[G{\"u}rkan et~al., 1999]{gurkan1999sample}
G{\"u}rkan, G., Yonca~{\"O}zge, A., and Robinson, S.~M. (1999).
\newblock Sample-path solution of stochastic variational inequalities.
\newblock {\em Mathematical Programming}, 84(2):313--333.

\bibitem[Habis and Csercsik, 2014]{habis2014cooperation}
Habis, H. and Csercsik, D. (2014).
\newblock Cooperation with externalities and uncertainty.
\newblock {\em Networks and Spatial Economics}, 15(1):1--16.

\bibitem[Hubert and Cobanli, 2015]{hubert2015pipeline}
Hubert, F. and Cobanli, O. (2015).
\newblock Pipeline power: a case study of strategic network investments.
\newblock {\em Review of Network Economics}, 14(2):75--110.

\bibitem[Hubert and Ikonnikova, 2011]{hubert2011investment}
Hubert, F. and Ikonnikova, S. (2011).
\newblock Investment options and bargaining power in the {Eurasian} supply
  chain for natural gas.
\newblock {\em The Journal of Industrial Economics}, 59(1):85--116.

\bibitem[Hubert and Orlova, 2018]{hubert2014access}
Hubert, F. and Orlova, E. (2018).
\newblock Network access and market power.
\newblock {\em Energy Economics}, 76:170--185.

\bibitem[Hubert and Suleymanova, 2008]{hubert2008dynamic}
Hubert, F. and Suleymanova, I. (2008).
\newblock Strategic investment in international gas-transport systems: {A}
  dynamic analysis of the hold-up problem.
\newblock {\em DIW Discussion Paper}, 846.

\bibitem[K{\'o}czy, 2007]{Koczy_recursive}
K{\'o}czy, L.~{\'A}. (2007).
\newblock A recursive core for partition function form games.
\newblock {\em Theory and Decision}, 63(1):41--51.

\bibitem[K\'{o}czy, 2018]{Koczy2018}
K\'{o}czy, L.~A. (2018).
\newblock {\em Partition function form games: {Coalitional} games with
  externalities}.
\newblock Number~48 in Theory and Decision Library C. Springer.

\bibitem[McQuillin, 2009]{mcquillin2009extended}
McQuillin, B. (2009).
\newblock The extended and generalized {Shapley} value: {Simultaneous}
  consideration of coalitional externalities and coalitional structure.
\newblock {\em Journal of Economic Theory}, 144(2):696--721.

\bibitem[Montero, 2005]{Montero2005}
Montero, M. (2005).
\newblock On the nucleolus as a power index.
\newblock {\em Homo Oeconomicus}, 22(4):551--567.

\bibitem[Roson and Hubert, 2015]{roson2015bargaining}
Roson, R. and Hubert, F. (2015).
\newblock Bargaining power and value sharing in distribution networks: a
  cooperative game theory approach.
\newblock {\em Networks and Spatial Economics}, 15(1):71--87.

\bibitem[Roth, 1988]{roth1988shapley}
Roth, A.~E. (1988).
\newblock {\em {The Shapley value: essays in honor of Lloyd S. Shapley}}.
\newblock Cambridge University Press.

\bibitem[Schmeidler, 1969]{Schmeidler1969}
Schmeidler, D. (1969).
\newblock The nucleolus of a characteristic function game.
\newblock {\em SIAM Journal of Applied Mathematics}, 17(6):1163--1170.

\bibitem[Shapley, 1953]{Shapley1953}
Shapley, L.~S. (1953).
\newblock A value for $n$-person games.
\newblock {\em Annals of Mathematics Studies}, 28:307--318.

\bibitem[Shenoy, 1979]{Shenoy1979}
Shenoy, P.~P. (1979).
\newblock On coalition formation: {A} game-theoretical approach.
\newblock {\em International Journal of Game Theory}, 8(3):133--164.

\bibitem[Sziklai et~al., 2018]{Sziklai2018}
Sziklai, B., K\'{o}czy, L.~A., and Csercsik, D. (2018).
\newblock The geopolitical impact of {Nord Stream 2}.
\newblock IEHAS Discussion Papers MT-DP 2018/21, Institute of Economics, Centre
  for Economic and Regional Studies, Hungarian Academy of Sciences.

\bibitem[Thrall and Lucas, 1963]{Thrall:1963}
Thrall, R. and Lucas, W. (1963).
\newblock $n$-person games in partition function form.
\newblock {\em Naval Research Logistics Quarterly}, 10:281--298.

\bibitem[Weiner, 2016]{weiner2016central}
Weiner, C. (2016).
\newblock Central and {East European} diversification under new gas market
  conditions.
\newblock Working Papers 2016/221, Institute of World Economics, Centre for
  Economic and Regional Studies, Hungarian Academy of Sciences.

\bibitem[Young, 1985]{Young1985}
Young, H. (1985).
\newblock Monotonic solutions of cooperative games.
\newblock {\em International Journal of Game Theory}, 14:65--72.

\end{thebibliography}

\end{document}